\documentstyle[12pt]{article}
\catcode`\@=11
\def\makepreprititle{\par
  \begingroup
  \def\thefootnote{\fnsymbol{footnote}}
  \def\
@makefnmark{\hbox
  to 0pt{$^{\@thefnmark}$\hss}}
  \if@twocolumn
  \twocolumn[\@makepreprititle]
  \else \newpage
  \global\@topnum\z@
  \@makepreprititle \fi\thispagestyle{empty}\@thanks
  \endgroup
  \setcounter{footnote}{0}
  \let\makepreprititle\relax
  \let\@makepreprititle\relax
  \gdef\@thanks{}\gdef\@author{}\gdef\@title{}
  \gdef\@preprintnumber{}\gdef\@preprintdate{}\gdef\subtitle{}
  \let\thanks\relax}
\def\preprintnumber#1{\gdef\@preprintnumber{#1}}
\def\preprintdate#1{\gdef\@preprintdate{#1}}
\def\subtitle#1{\gdef\@subtitle{#1}}
\def\@makepreprititle{\newpage
{\def\baselinestretch{1}
  \begin{flushright} \@preprintnumber \par
  \@preprintdate \end{flushright} } \par
  \begin{center}
\vskip 1.5em
  {\LARGE \@title \par} \vskip 2.5em
  {\large \lineskip .5em
  \begin{tabular}[t]{c}\@author
  \end{tabular}\par}
  \vskip 1em {\large \@date} \end{center}
  \par
  \vfil}
\date{\sl Department of Physics, Tohoku University\\Sendai, 980 Japan}
\preprintdate{~}
\preprintnumber{~}
\subtitle{}
\def\endabstract{\if@twocolumn\else\endquotation\fi}
\catcode`\@=12

\preprintnumber{DPSU-9402\\LBL-35731\\UCB-PTH-94/16\\TU-459}
\preprintdate{June 1994}
\title{\Large
Low-Energy Effective Lagrangian\\
in Unified Theories\\
with Non-Universal Supersymmetry Breaking Terms\thanks{This work was
supported in part by the Director, Office of Energy Research, Office
of High Energy and Nuclear Physics, Division of High Energy Physics of
the U.S. Department of Energy under Contract DE-AC03-76SF00098.}}
\author{Yoshiharu Kawamura,$^1$  Hitoshi Murayama$^{2,3}$
\\  and \\ Masahiro Yamaguchi$^3$}
\date{\sl $^1$ Department of Physics, Shinshu University\\
Matsumoto, 390 Japan\\
$^{2}$ Theoretical Physics Group, Lawrence Berkeley Laboratory \\
University of California, CA 94720\\
$^3$ Department of Physics, Tohoku University\\
Sendai, 980-77 Japan}

\newcommand{\cleqn}{\setcounter{equation}{0}}
\def\lsim{\mathrel{\mathpalette\vereq<}}

\catcode`\@=11
\def\vereq#1#2{\lower3pt\vbox{\baselineskip1.5pt \lineskip1.5pt
\ialign{$\m@th#1\hfill##\hfil$\crcr#2\crcr\sim\crcr}}}
\catcode`\@=12

\begin{document}

\makepreprititle
\begin{abstract}
Supersymmetric grand unified theories with non-universal soft
supersymmetry breaking terms are studied.  By integrating out the
superheavy fields at an unification scale, we compute their low-energy
effective Lagrangian.  We find new contributions to the scalar
potential specific to the non-universal supersymmetry breaking.
$D$-term contribution to the scalar masses is one example.  The gauge
hierarchy achieved by a fine-tuning in the superpotential would be
violated in general due to the non-universal SUSY breaking terms.  We
show, however, it is preserved for a certain class of the soft terms
derived from a {\em hidden} ansatz.  We also discuss some
phenomenological implications of the non-universal supersymmetry
breaking, including predictions of the radiative electroweak symmetry
breaking scenario and of  no-scale type models.
\end{abstract}

\newpage

\section{Introduction}

Supersymmetry (SUSY) has been regarded as a beautiful mechanism which
ensures the stability of the hierarchy,  between the weak scale and the
grand-unification (GUT)  \cite{SUSY-GUT} or Planck scale against radiative
corrections.  There is, however, another virtue of SUSY which is not
often quoted: it is that the SUSY breaking terms at the weak scale
directly reflect the physics at very high energies thanks to the
moderate renormalization effects, {\it i.e.}, the absence of the quadratic
divergences. For instance, one can study experimentally whether the
weak scale SUSY spectrum is consistent with GUT, by
measuring the gaugino masses \cite{JLC}. Also, the scalar mass
spectrum has certain ``sum rules'' specific to symmetry breaking patterns
\cite{KMY,KT}.  Therefore, it is important to know the predictions of the
various models on the SUSY breaking terms at low energy.

Many efforts have been devoted to the low-energy predictions under the
assumption of the universal SUSY breaking terms, which led to remarkable
progress in recent years \cite{Kane,Nanopoulos}. The idea
of the universal SUSY breaking terms was led by the non-observation of
the large flavor changing neutral current processes due to the SUSY
particle loops \cite{FCNC}. And the hidden sector SUSY breaking
\cite{Arnowitt} combined with the minimal supergravity indeed
leads to the
universal SUSY breaking terms, with or without grand unification
\cite{HLW}.
However a decade after these works, there are increasing interests in the
non-universal form of the SUSY breaking terms, at least due to the
following two reasons. (1) The unification scale is now believed to be
substantially lower than the gravitational scale
$M_{Pl}/\sqrt{8\pi}$,\footnote{Note that the difference between the
GUT scale $\simeq 10^{16}$~GeV and the gravitational scale $\simeq
10^{18}$~GeV is one seventh of that between the GUT scale and the weak scale
if measured in log-scale.}
and the radiative correction changes the form of the supersymmetry
breaking terms. (2) The superstring theory implies highly non-universal
form of the K\"ahler potential in general \cite{string}.
Therefore, it is an important task to see what
low-energy theory results from the unified theories with
non-universal form of the supersymmetry breaking terms.

There are some indications that the non-minimal SUSY breaking terms give
potentially important consequences in the low-energy effective
Lagrangian. First, they give rise to the so-called $D$-term contribution
to the scalar masses when the rank of the gauge group is reduced
\cite{DH}.  We pointed out in the previous letter \cite{KMY} that they
can be generated even at the GUT scale and give observable
consequences on the weak scale scalar mass spectrum which is important
to distinguish the various symmetry breaking patterns \cite{KMY,KT}.
Second, the non-universal SUSY breaking terms may ruin the fine-tuning
in SUSY-GUT models that the Higgs doublet may acquire a mass of
intermediate scale in general. Third, the non-minimal initial
conditions at the GUT scale modify the phenomenological conclusions
made in the literature, in predictions on neutralino cosmic abundance,
radiative breaking scenario and so on.

In this paper, we derive the low-energy effective Lagrangian starting
 from the grand unified theories with non-universal SUSY breaking terms.
In fact, we find new contributions to the SUSY breaking terms at
the low energy, after integrating out the superheavy fields at the
GUT scale. It of course reduces to that obtained in Ref.~\cite{HLW} in
the case with the universal SUSY breaking terms.
Furthermore, the non-minimality of the SUSY breaking terms at the GUT
scale can lead to many interesting phenomenological consequences.  We
point out, for instance, that the scalar lepton becomes always heavier
than the bino due to the radiative correction between Planck and GUT
scales, and the upper bound on the slepton mass in the no-scale type models
\cite{IKYY,LNY}
becomes invalid.   Possible effects on radiative breaking scenario are
also discussed.

The paper is organized as follows.  In section 2, we first review the
low-energy Lagrangian with universal SUSY breaking terms. Then we point
out the importance of studying the case of the non-universal SUSY
breaking terms,  and
demonstrate its important outcomes in explicit examples. We
give the low-energy effective Lagrangian with the non-universal SUSY
breaking terms in section 3. Here we also give an ansatz of the SUSY
breaking terms based on the assumption that the hidden sector is hidden;
this ansatz is stable under the renormalization effects,
and we show that the fine-tuning in SUSY-GUT is not ruined in this ansatz
of the SUSY breaking terms. We point out phenomenological implications
of the non-universal SUSY breaking terms in section 4.
Section 5 is devoted to conclusions.

\section{Why Non-universal Soft SUSY Breaking Terms?}
\cleqn
In this section, we briefly explain the typical effects of the
non-universal soft SUSY breaking terms, to demonstrate the potential
importance of their low-energy consequences. First we review the basic
results by Hall et al. \cite{HLW}, where the low-energy
effective Lagrangian is derived in the case with the universal SUSY breaking
terms. We then point
out that the universal form
of the SUSY breaking terms is not preserved by the radiative
corrections. As examples of potential importance of the non-universal
SUSY breaking terms, we show that the non-universal scalar mass of the
superheavy fields gives rise to $D$-term contributions to the scalar
masses in the low-energy effective Lagrangian. We also give an example
that the non-universal soft SUSY breaking terms destabilizes the
hierarchy between the unification and weak scales. These
observations give us a motivation to study the low-energy consequence
of the non-universal SUSY breaking terms in a more general framework.

\subsection{Minimal Supergravity}

The low-energy effective Lagrangian of the minimal supergravity was shown
to be extremely simple in Ref.~\cite{HLW}. Though the discussion in that
paper was based on the supergravity Lagrangian, we re-phrase their
result in the flat limit, {\it i.e.}\/, in the context of
the global SUSY Lagrangian
with soft SUSY breaking terms.\footnote{This treatment can be
actually justified from their analysis, that the fields in the hidden
sector do not shift as the light fields are varied at $O(m_{S})$.}

The minimal supergravity suggests the following form of the scalar
potential in the observable sector,
\begin{eqnarray}
V &=& V_{SUSY} + V_{
        \begin{picture}(25,0)(0,0)
        \put(0,0){\scriptsize $SUSY$}
        \put(0,0){\line(4,1){22}}
        \end{picture} } ,\\
V_{SUSY} &=& - F^\kappa F_\kappa^* - \frac{1}{2} D^\alpha D^\alpha
	+ F^\kappa \frac{\partial W}{\partial z^\kappa} + {\it h.c.}
	+ g_\alpha D^\alpha z_\kappa^* (T^\alpha)^\kappa_\lambda
		z^\lambda,\\
V_{
        \begin{picture}(25,0)(0,0)
        \put(0,0){\scriptsize $SUSY$}
        \put(0,0){\line(4,1){22}}
        \end{picture} }
	&=& A W + B z^\kappa \frac{\partial W}{\partial z^\kappa}
		+ {\it h.c.}
	+ |B|^2 z_\kappa^* z^\kappa,
\end{eqnarray}
where $W$ is the superpotential, $z^\kappa$ are the scalar fields, and
$A$, $B$ are soft SUSY breaking parameters.\footnote{The
definition here is related to that in Ref.~\cite{HLW} by $m_g = B^*$ and
$m'_g = (A+3B)^*$. It is noteworthy that they also discussed a modification
of the minimal supergravity based on $U(n)$ invariance in the kinetic
function to ensure the universality of the scalar masses.} This form of
the SUSY breaking
terms is referred to as ``universal,'' because all the scalar masses are
equal.

The main result in Ref.~\cite{HLW} is that the following form of the
SUSY breaking terms results in the low-energy Lagrangian after
integrating out the superheavy fields in the above Lagrangian,
\begin{equation}
V_{
        \begin{picture}(25,0)(0,0)
        \put(0,0){\scriptsize $SUSY$}
        \put(0,0){\line(4,1){22}}
        \end{picture} }
	= -2A W_{\it eff}
	+ (A+B) z^k \frac{\partial W_{\it eff}}{\partial z^k }
		+ {\it h.c.}
	+ |B|^2 z_k^* z^k,
\end{equation}
and it still has the same form as the original one by suitable
redefinition of the $A$ and $B$ parameters except the mass squared terms.
Here $z^k$ are the light scalar fields and
$W_{\it eff}$ is
the superpotential $W$ with the extremum values for superheavy fields
plugged in.
Moreover, the scalar mass terms
are still universal with the same mass, $B$. It is noteworthy that
the authors of Ref.~\cite{HLW} did the analysis including the hidden
sector fields, and proved
that they do not shift when the light fields fluctuate at $O(m_{S})$
where $m_{S}$ is the SUSY breaking scale  $\sim 1$ TeV,
while the constant term in the superpotential should be shifted to
cancel the cosmological constant.

The remarkable simplicity of the low-energy Lagrangian led to a number
of strong conclusions, like the natural absence of the flavor changing
neutral currents \cite{FCNC} or radiative breaking scenario due to the
heavy top quark \cite{radiative-breaking}. Due to these successes, it
became like a dogma in the phenomenological analysis of the SUSY
models. However, it becomes increasingly apparent that the
supergravity Lagrangian may not have the minimal form, and it may lead
to important consequences on the low-energy effective Lagrangian, as
will be discussed in the next subsections.

\subsection{Naturalness of the Universal SUSY Breaking Terms}
\label{subsec:naturalness}
It was pointed out that the higher order corrections in general
destroy the minimal form of the K\"ahler potential
\cite{Russian,Gaillard}. This poses a question on the naturalness of
the minimal supergravity Lagrangian. However, this discussion is
based on the one-loop corrections with a naive cut-off set at the Planck
scale, and such higher order corrections may be absent in specific
dynamics beyond the Planck scale. Note that even when the Planck scale
dynamics satisfies certain symmetry to ensure the minimal form of the
SUSY breaking terms, we still expect that it will be modified by the
radiative corrections below the Planck scale.

Let us give an example of the minimal $SU(5)$ model
with vanishing scalar masses at the Planck
scale \cite{no-scale}. The scalar fields acquire their masses
through renormalization group equations, which are different for $
{\bf \bar
5}$ and ${\bf 10}$ representations. Their masses at the $SU(5)$
breaking scale $\simeq 2\times10^{16}$~GeV are
\begin{eqnarray}
m_{\bar  5}^2 &=& 0.30 M^2\\
m_{10}^2 &=& 0.45 M^2,
\end{eqnarray}
where $M$ refers to the $SU(5)$ gaugino mass at the supergravity scale
$M_{Pl}/\sqrt{8\pi}$. Apparently these contributions cannot be
neglected in the phenomenological analyses, and are also non-universal.
They become even larger in the non-minimal GUT models because the gauge
coupling constant tends to be larger than that in the minimal $SU(5)$
model. This demonstrates that the running between the Planck scale and
the $SU(5)$ breaking scale is not negligible \cite{MoxhayYamamoto}.
Then the SUSY breaking terms do not have a minimal form
any more at the GUT scale, where the superheavy fields should be
integrated out.

Instead of the universal SUSY breaking terms, we will take the following
ansatz for the SUSY breaking terms
\begin{equation}
V_{
        \begin{picture}(25,0)(0,0)
        \put(0,0){\scriptsize $SUSY$}
        \put(0,0){\line(4,1){22}}
        \end{picture} }
	=m_S A W +m_S B^\kappa (z)
    \frac{\partial W}{\partial z^\kappa}+h.c.
	+ O(m_S^2) \ {\rm terms}.
\label{special}
\end{equation}
(See section \ref{subsub:softSUSY-breaking-terms} for the
notation.)\footnote{$B^\kappa$ are holomorphic functions of $z^\lambda$
independent for each $\kappa$, and {\it not}\/ the derivative of a
single function $B$.}
It will be shown that this form of the SUSY breaking terms is stable
under renormalization, or in other words, natural in the weak sense.
Since the minimal form is not stable under the renormalization, we
believe this is the framework to work out the low-energy effective
Lagrangian. Moreover, from the view point of supergravity,
this form is the most general form of the SUSY
breaking terms induced by superhiggs mechanism
with the assumption that the hidden sector is
`hidden', {\it i.e.}, that the superpotential is a sum of two
independent pieces consisting of observable and hidden fields,
respectively.  We leave the detail to
section~\ref{subsub:softSUSY-breaking-terms}.

Recall, also, that the superstring theory suggests non-minimal forms of the
SUSY breaking terms in general \cite{stringII}. This is especially
true when the SUSY is broken by the $F$-component of the moduli
fields. The scalar masses depend on their modular weights.  The
gaugino masses are also non-universal.

These observations give us a strong motivation to study the low-energy
effective Lagrangian from the unified theories with non-universal SUSY
breaking terms.  In the next two subsections,  we will point out potentially
important effects of the non-universality.

One remark on the squark degeneracy is in order. It has been often
stated that one needs high degeneracy of the scalar masses to ensure the
potentially large contribution of the squark loop diagrams to the
$K^0$-$\overline{K}^0$ mixing. However, this does not require the
degeneracy of all scalar masses at the GUT or Planck scale. The only
requirement is  the degeneracy of the first- and second-generation
squarks with the same quantum numbers. Note that neither the $D$-term
contributions nor the renormalization group evolution due to the gauge
interactions destroy the degeneracy as far as they do not distinguish
the generations of the light quarks. One may have highly degenerate
squark masses due to the gluino mass contribution even with the
different masses as the initial conditions at the Planck scale. Though
the flavor changing neutral current still puts strong constraint on the
non-minimality, it does not diminish our interest to study its
low-energy consequences.

\subsection{Possible $D$-term Contributions}

Though the low-energy Lagrangian is surprisingly simple when it had
universal form at the unification scale, there arise different
contributions to $V_{
        \begin{picture}(25,0)(0,0)
        \put(0,0){\scriptsize $SUSY$}
        \put(0,0){\line(4,1){22}}
        \end{picture} }$ when one integrates out the superheavy fields
 from the Lagrangian with non-universal form of the SUSY breaking terms. This
has an important consequence on the scalar masses when one probes the
symmetry breaking pattern from the weak-scale measurements of the masses
\cite{KMY,KT} (see section 4.1). In fact, we pointed out that the
$D$-term contribution may arise at the GUT-scale \cite{KMY}, and there have
appeared papers which discuss the effect of the $D$-term contributions to the
Higgs masses in the radiative breaking scenario \cite{Hempfling,HRS}.

In this subsection, we
give a simple example which generates a $D$-term contribution to the
low-energy scalar masses.
Take a simple
superpotential $W = h(\phi_1 \phi_2 - \mu^2) \chi$, with $U(1)$-charges
$Q=+1$ for $\phi_1$, $Q=-1$ for $\phi_2$ and $Q=0$ for $\chi$.
We also have light fields
$z^k$ which couple to the $U(1)$ gauge fields but do not couple
to the heavy fields $\phi_1$, $\phi_2$ and $\chi$ in the
superpotential. Then the full potential
reads as
%\begin{eqnarray}
%V &=& h^2 ( |\phi_1|^2 + |\phi_2|^2 ) |\chi|^2
%	+ h^2 |\phi_1 \phi_2 - \mu^2 |^2
%	+ \frac{g^2}{2} (|\phi_1|^2 -  |\phi_2|^2
%		+ \sum_k Q_k |z^k|^2)^2 \nonumber \\
%& &	+ h A (\phi_1 \phi_2 \chi + \phi_1^* \phi_2^* \chi^*)
%	+ m_1^2 |\phi_1|^2 + m_2^2 |\phi_2|^2 + m_\chi^2 |\chi|^2,
%\end{eqnarray}
\begin{eqnarray}
V &=& - |F_1|^2 - |F_2|^2 - |F_\chi|^2 - |F^k|^2 - \frac{1}{2} D^2
	\nonumber \\
& & + (F_1 h \phi_2 \chi + F_2 h \phi_1 \chi
	+ F_\chi h (\phi_1 \phi_2 - \mu^2) + h.c.) \nonumber \\
& & + g D \left(|\phi_1|^2 - |\phi_2|^2 + \sum_k Q_k |z^k|^2\right)
	\nonumber \\
& & + (A h \phi_1 \phi_2 \chi - C h \mu^2 \chi + h.c.) \nonumber \\
& & + m_1^2 |\phi_1|^2 + m_2^2 |\phi_2|^2 + m_\chi^2 |\chi|^2 + m_k^2 |z^k|^2,
\end{eqnarray}
where $F_1$, $F_2$, $F_\chi$, $F^k$ are
the auxiliary fields of $\phi_1$, $\phi_2$, $\chi$, $z^k$, respectively,
and $D$ is the auxiliary field of the $U(1)$ gauge field.\footnote{One
can also study the same potential after integrating out the auxiliary
fields, as one usually does. However we keep the auxiliary fields as
independent variables for the later convenience.} The SUSY breaking terms
$A$, $C$, $m_{\kappa}^2$ are $O(m_{S})$, $O(m_{S})$, $O(m_{S}^2)$,
respectively.

We integrate out the heavy fields $\phi_1$, $\phi_2$ and $\chi$ from this
potential, taking light fields $z^k$ fluctuating at $O(m_{S})$. The
stationary solutions to the heavy fields are solved by expanding
in powers of $m_{S}/\mu$, and one finds
\begin{eqnarray}
\lefteqn{\phi_1 = \mu} \nonumber \\
&+& \!\!\!\! \left[
	\frac{-A^2+C^2}{8 h^2 \mu} - \frac{m_1^2 + m_2^2}{4 h^2 \mu}
	- \frac{m_1^2 - m_2^2}{8 g^2 \mu}
	- \frac{\sum_k Q_k |z^k|^2}{4g\mu} \right]
%\nonumber \\
%	&+&
+ O\left( \frac{m_{S}^3}{\mu^2} \right), \\
\lefteqn{\phi_2 = \mu} \nonumber \\
&+& \!\!\!\! \left[
	\frac{-A^2+C^2}{8 h^2 \mu} - \frac{m_1^2 + m_2^2}{4 h^2 \mu}
	+ \frac{m_1^2 - m_2^2}{8 g^2 \mu}
	+ \frac{\sum_k Q_k |z^k|^2}{4g\mu} \right]
%\nonumber \\
%	&+&
+ O\left( \frac{m_{S}^3}{\mu^2} \right), \\
\lefteqn{\chi = -\frac{A-C}{2h} + O\left( \frac{m_{S}^3}{\mu^2} \right),}
\end{eqnarray}
and for the auxiliary fields,
\begin{eqnarray}
F_1 &=& h \phi_2 \chi = -\frac{(A-C)}{2} \mu
	+ O\left( \frac{m_{S}^3}{\mu} \right) , \\
F_2 &=& h \phi_1 \chi = -\frac{(A-C)}{2} \mu
	+ O\left( \frac{m_{S}^3}{\mu} \right) , \\
F_\chi &=& h(\phi_1 \phi_2 - \mu^2)
	= \frac{-A^2+C^2}{4h} - \frac{m_1^2+m_2^2}{2h}
	+ O\left( \frac{m_{S}^3}{\mu} \right) , \\
D &=& g \left(|\phi_1|^2 - |\phi_2|^2 + \sum_k Q_k |z^k|^2 \right)
	= -\frac{m_1^2 - m_2^2}{2g} + O\left( \frac{m_{S}^3}{\mu} \right) .
\end{eqnarray}
The low-energy effective Lagrangian can be obtained by plugging these
solutions into the original Lagrangian, giving
\begin{eqnarray}
V_{\it eff} &=&
	(\mbox{$z^k$-independent terms of $O(m_{S}^2 \mu^2)$})
		\nonumber \\
& &	+ m_k^2 |z^k|^2
	- \frac{1}{2} (m_1^2 - m_2^2) \sum_i Q_k |z^k|^2
		\nonumber \\
& &	+ (\mbox{terms of $O(m_{S}^5/\mu)$}).
\end{eqnarray}
The latter terms are the contributions from the non-vanishing value of the
$D$-term, giving rise to different masses for different quantum numbers.
One also sees that the $D$-term contributions vanish if the scalar
masses are universal, consistent with the analysis in Ref.~\cite{HLW}
reviewed in the previous subsection.

The $D$-term contribution exists in general when the rank of the gauge
group is reduced by the symmetry breaking \cite{DH}, and the SUSY breaking
terms are non-universal. They give rise to observable effects at the
weak scale.

\subsection{Instability of the Hierarchy}

The non-universal SUSY breaking terms have a dramatic consequence when
there is a fine-tuning to keep light fields at the weak scale. The light
fields acquire masses of the order of the intermediate scale in general.
It seems to us that %the existence of the dangerous term (\ref{mixingA})
this problem is not widely recognized in the literature (see
however \cite{MoxhayYamamoto,Ibanez}).

In the SUSY standard model,
we need (at least) two Higgs doublets with opposite hypercharges and
these doublet scalars form a SUSY-breaking mixing mass term.
Indeed the problem we now discuss  is related to
the fine-tuning problem of the Higgs doublet mass (the gauge hierarchy
problem), which is inevitable
in a wide class of SUSY GUT models in order to obtain the light Higgs
doublets of the SUSY standard model.  We will explain the problem in
some detail. The discussions given below are based on  the
observation in an unpublished work \cite{IMMOYY}.

To exemplify the problem, let us consider the minimal SUSY $SU(5)$
model \cite{SUSY-GUT} whose superpotential is
\begin{equation}
W = \lambda {\rm tr} \Sigma^3 + M_\Sigma {\rm tr} \Sigma^2
	+ H_u (f \Sigma + M_H) H_d.
\end{equation}
Here $\Sigma$, $H_u$ and $H_d$ are the fields of {\bf 24}, {\bf 5} and
${\bf \bar 5}$
representations of $SU(5)$, respectively. $\lambda$, $f$ are
dimensionless coupling constants, while $M_\Sigma$, $M_H$ are GUT scale
mass parameters.
If we add the SUSY breaking terms arbitrarily, the potential reads
\begin{eqnarray}
V &=& \left| \frac{\partial W}{\partial \Sigma} \right|^2
	+ \left| \frac{\partial W}{\partial H_u} \right|^2
	+ \left| \frac{\partial W}{\partial H_d} \right|^2
	+ g^2 \sum_\alpha \left(\Sigma^\dagger [T^\alpha, \Sigma]
		+ H_u^\dagger T^\alpha H_u
		+ H_d^\dagger (-T^{\alpha*}) H_d \right)^2	\nonumber\\
& &+	\{
		\lambda A_\Sigma {\rm tr} \Sigma^3
		+ B_\Sigma M_\Sigma {\rm tr} \Sigma^2
		+ f A_H H_u \Sigma H_d + B_H M_H H_u H_d
\nonumber \\
& & + h.c.
	\}, \label{SUSYbreakingSU5}
\end{eqnarray}
where $A_\Sigma$, $B_\Sigma$, $A_H$ and $B_H$ are the SUSY breaking
parameters of order $m_{S}$ and we have omitted the SUSY breaking
scalar masses which are irrelevant to the following discussions.
The $SU(5)$ gauge coupling constant is denoted by $g$. Taking
\begin{equation}
\Sigma = \frac{1}{\sqrt{60}} \left(
	\begin{array}{ccccc}
	2&&&&\\
	&2&&&\\
	&&2&&\\
	&&&-3&\\
	&&&&-3
	\end{array}
	\right)
	\sigma,
\end{equation}
the minimal of the potential is located at $
\sigma_0 = \frac{2\sqrt{60}}{3\lambda} M_\Sigma $ in the SUSY limit, which is
shifted by $\delta \sigma= \frac{\sqrt{60}}{3\lambda}
(A_{\Sigma}-B_{\Sigma})$ in the presence of the SUSY breaking terms.
The mixing mass of
the two doublet Higgs bosons $m_{12}^2H_u H_d$ %\footnote{Hereafter
%$H_u$ and $H_{d}$ represent the $SU(2)_L$ doublet Higgs multiplets.}
is given by
\begin{equation}
m_{12}^2 =  \frac{3f}{\sqrt{60}} \sigma_0
		(A_\Sigma - B_\Sigma - A_H + B_H)+O(m_{S}^2),
\end{equation}
where we have used that the supersymmetric mass of the  Higgs
doublets is fine-tuned to be
\begin{equation}
  M_{H}-\frac{3g}{\sqrt{60}}\sigma_0 =O(m_{S})
\end{equation}
at the SUSY limit.  Clearly for a class of the SUSY breaking
parameters where the combination $A_\Sigma - B_\Sigma - A_H + B_H $ does
not vanish,
$m_{12}^2$ lies at an intermediate scale $\sim m_{S} M_X$ and the
gauge hierarchy is destabilized.

A crucial observation is, however, that the mixing mass $m^2_{12}$
becomes of order $m_{S}^2$ if we assume the form of the soft terms as in
Eq.~(\ref{special}).  Indeed for the special form of the SUSY breaking terms
\begin{equation}
   AW +B_1 \Sigma \frac{\partial W}{\partial \Sigma }
        +B_2 H_u \frac{\partial W}{\partial H_u}
        +B_3 H_d \frac{\partial W}{\partial H_d},
\end{equation}
which follows from the ansatz in section \ref{subsec:naturalness},
the coefficients $A_{\Sigma}$ {\it etc.} in Eq.~(\ref{SUSYbreakingSU5})
are written as
\begin{eqnarray}
   & A_{\Sigma}=& A+ 3 B_1, \\
   & B_{\Sigma}=& A+ 2 B_1, \\
   & A_H =& A +B_1+B_2+B_3, \\
   & B_H =& A +B_2+B_3.
\end{eqnarray}
Then we find
\begin{equation}
A_\Sigma - B_\Sigma - A_H + B_H = 0,
\end{equation}
which guarantees the lightness of the Higgs doublets in the minimal
$SU(5)$ model.

\section{Scalar Potential in the Effective Theory}
\cleqn
In the previous section, we have examined some examples to demonstrate
that the non-universal SUSY breaking terms can lead to important
consequences.  In this section, we will give a more general discussion.
We consider a softly-broken supersymmetric unified theory\footnote{The
gauge group of the theory is not necessarily grand-unified into a
simple group.} whose gauge group is broken at an energy scale $M_X$.
We will derive the scalar potential $V_{\it eff}$ in the low-energy
effective theory by integrating out the heavy sector.

\subsection{General Discussion}
\label{subsec:general-argument}

We first list the basic assumptions in the following discussion.
\begin{enumerate}
\item The unified theory is described as a renormalizable
supersymmetric theory with soft SUSY breaking terms.
\item The unified gauge symmetry is broken at a scale $M_X$
 which is much higher than the SUSY breaking scale $m_{S}$ (of
$O(1)$ TeV).
\item SUSY is not spontaneously broken in the absence of the soft SUSY
breaking terms in the Lagrangian.
\item All the particles can be classified as heavy (with mass $O(M_X)$)
or light (with mass $O(m_{S})$) in the absence of the SUSY breaking
terms.
\item The light scalar fields have vacuum expectation values
(VEVs)
 as well as fluctuations only of $O(m_{S})$.
\end{enumerate}
The assumptions here are basically the same as in Ref.~\cite{HLW}.

We solve the stationary conditions of the potential of the full theory
for the heavy scalar fields while keeping the light scalar fields
arbitrary.  We then ``integrate out'' the heavy fields by inserting
the solutions of the stationary conditions into the potential.
The potential obtained in this way should be regarded as the potential of
the low-energy effective theory renormalized at the scale $M_X$.
In deriving the effective scalar potential, we fully utilize
 the equalities from the gauge invariance
given in Appendix A.

The procedure to obtain the low-energy effective potential in this
paper is quite similar to that of Ref.~\cite{HLW}, though they
started from the (minimal) supergravity Lagrangian whereas our starting
point is the softly-broken global SUSY.  As for the SUSY
breaking, we consider a non-universal form which is expected to be
realized in realistic models as we
discussed in the last section.

%(1) The supergravity Lagrangian itself may have
%non-minimal K\"ahler potential, and
%give rise to non-universal SUSY breaking terms after integrating out
%the hidden sector.
%(2) Even if the supergravity Lagrangian is minimal,
%the universality can be broken owing to
%the renormalization group running of the SUSY breaking terms
%from the gravitational scale down to the symmetry breaking scale $M_X$.

\subsubsection{SUSY Limit}
\label{subsubsec:SUSY-limit}
First we review the basic properties of the scalar potential in the
absence of the SUSY breaking terms.

The superpotential is a holomorphic function of chiral superfields $\Phi$
\begin{equation}
	W =  W (\Phi^\kappa).
\end{equation}
 The supersymmetric scalar potential is given as
\begin{equation}
V_{SUSY} = 	- \frac{1}{2} (D^\alpha)^2
	- F_\kappa^* F^\kappa
	+  D^\alpha
		(z^\dagger T^\alpha z)
	+ F_\kappa^\ast \frac{\partial W^*}{\partial z_\kappa^*}
	+ \frac{\partial W}{\partial z^\kappa} F^\kappa,
\label{SUSYpart}
\end{equation}
where $z^\kappa$ are the scalar components of the chiral superfields
$\Phi^\kappa$.  $F^\kappa$ is the auxiliary field of $\Phi^\kappa$,
$T^\alpha$ stands for a gauge generator and $D^\alpha$ is its $D$-term.
$z^\dagger T^\alpha z$ is an abbreviation of $ z_\lambda^\ast
(T^\alpha)^\lambda_\kappa z^\kappa$. Summation over $\alpha$,
$\kappa$ and $\lambda$ is implied.

The stationary conditions of the potential are simply
\begin{eqnarray}
F_\kappa^\ast = \frac{\partial W}{\partial z^\kappa} = 0,
	\label{SUSYF}\\
D^\alpha = z^\dagger T^\alpha z = 0
	\label{SUSYD}
\end{eqnarray}
under the condition that SUSY is unbroken.
We denote the solutions to the stationary conditions as $z^\kappa =
z_0^\kappa$.

We can always take a basis of $z^\kappa$ to diagonalize the
fermion mass matrix
\begin{equation}
\mu_{\kappa\lambda} = \left.
	\frac{\partial^2 W}{\partial z^\kappa \partial z^\lambda}
		\right|_{z = z_0} .
\end{equation}
Then the scalar fields are classified \cite{HLW} either as ``heavy''
fields $z^K, z^L,
\cdots$,  ``light'' fields $z^k, z^l, \cdots$, or Nambu--Goldstone fields
which will be discussed just below.

%$z^\alpha, z^\beta, \cdots$ which will be absorbed into the gauge
%multiplet.\footnote{To be more precise, either real or imaginary part of
%the $z^\alpha$ are the true Nambu--Goldstone bosons which are absorbed
%into the gauge bosons, and the other part acquire the same mass as that
%of the gauge bosons from the $D$-term in the potential.}

The mass matrix of the gauge bosons is
\begin{equation}
(M_V^2)_{\alpha\beta} =
	z_{0\lambda}^\ast \{ T^\alpha, T^\beta \}^\lambda_\kappa z_0^\kappa
	= 2  (z_0^\dagger T^\alpha)_\kappa
		(T^\beta z_0)^\kappa,
%	= g_\alpha g_\beta z^{\alpha\ast} z^\beta ,
\end{equation}
where we used Eq.~(\ref{SUSYD}) in the last equality, and it can be
diagonalized so that the gauge generators are classified into ``heavy''
(those broken at $M_X$) $T^A, T^B, \cdots$ and ``light'' (which remain
unbroken above $m_{S}$) $T^a, T^b, \cdots$. For  the heavy
generators, the fields $(T^A z_0)^\kappa$ correspond to the direction of
the Nambu--Goldstone fields in the field space, which span a vector
space with the same dimension as the number of heavy generators. We
can take a basis of the Nambu--Goldstone multiplets, $z^A, z^B,
\cdots$ so that
\begin{equation}
     \sqrt{2} (T^A z_0)^B = M_V^{AB} .
%\mu _A \delta^{AB}.
\end{equation}
Here the Nambu--Goldstone fields are taken to be orthogonal to the heavy and
light fields such as $(T^A z_0)^K = 0$, $(T^A z_0)^k =
0$. Also  the vanishing of the $D$-terms Eq.(\ref{SUSYD}) implies
\begin{equation}
   z^A_0=0.
\end{equation}

\subsubsection{ Soft SUSY Breaking Terms}
\label{subsub:softSUSY-breaking-terms}
The soft SUSY breaking terms can be classified by spurion insertions
as\footnote{We owe this classification to the discussions with K.~Inoue,
Y.~Okada and T.~Yanagida.}
\begin{eqnarray}
   & \int d^2 \theta (m_S \theta^2) U(\Phi)
 &=  m_S U(z),
\label{U} \\
   & \int d^4 \theta (m_S \theta^2) B(\Phi, \Phi^\dagger)
 &= m_S F_\lambda^* \frac{\partial B}{\partial z^*_\lambda},
\label{B} \\
   & \int d^4 \theta (m_S^2 \theta^2 \bar \theta^2)
    C(\Phi, \Phi^\dagger)
 &= m_S^2 C(z, z^*).
\label{C}
\end{eqnarray}
$U(\Phi)$ is a holomorphic function of the chiral superfields.  $B$ is
expressed as
\begin{equation}
     B(\Phi, \Phi^\dagger)=
       B_{2 \lambda}^\kappa \Phi^\dagger_\kappa \Phi^\lambda
\label{Bbilinear}
\end{equation}
for renormalizable theories and thus $\partial B/ \partial
z^*_\lambda$ is a function of $z$, not of $z^*$.  A term which depends
only on $\Phi^\dagger$ does not appear in Eq.~(\ref{Bbilinear}), since
it can be absorbed into the superpotential as\cite{GiudiceMasiero}
\begin{equation}
   \int d^4 \theta (m_S \theta^2) H(\Phi^\dagger)
  =\int d^2 \bar \theta m_S H(\Phi^\dagger).
\end{equation}
$C(\Phi, \Phi^\dagger)$ is a bilinear polynomial of  the chiral and
anti-chiral fields.

The non-renormalization theorem \cite{NR} implies that the form $U(z)$ is
preserved from radiative corrections since Eq.~(\ref{U}) is an
$F$-term.  On the other hand, the functions $B$ and $C$ are generally
renormalized as they are the $D$-terms.   For example, the minimal
supergravity induces the soft terms such as $U=A W(\Phi)$,
$B=\Phi^\dagger \Phi$ and $C=0$ in the flat limit with $A$ being a constant.
When we take radiative corrections into account,  $U$ remains the
same, but   $B$ and $C$ suffer from the renormalization and  in general become
non-universal.  This
observation leads us to investigate  non-universal soft terms.

In the rest of this paper, we take the ansatz
\begin{equation}
     U(\Phi) =A W (\Phi)   \label{ansatz}
\end{equation}
as in the case of the minimal supergravity, while we admit
non-universal structure for $B$ and $C$.  We can show that
Eq.~(\ref{ansatz}) is derived from non-minimal supergravity Lagrangian
in the flat limit, provided that the observed sector does not have
couplings in the superpotential to the sector which is responsible for
the spontaneous breaking of the local supersymmetry (the ``hidden''
sector) \cite{KL}.\footnote{Our usage of the term ``hidden sector'' is
somewhat different from that in Ref. \cite{KL}, where the moduli fields whose
$F$-components break SUSY are not included in the hidden sector.} On
phenomenological grounds this ``hidden'' assumption is widely
accepted, since otherwise the large SUSY breaking would directly be
transmitted to the observed sector and the SUSY is
badly broken in the low-energy effective theory as a consequence.  For
example, the
Yukawa coupling of the ``hidden'' sector field to the Higgs doublets
induces a large mass term of the Higgs bosons of order $m_S M_{Pl}$
and destabilizes the weak scale.
Furthermore we believe that the assumption allows us to integrate the
hidden sector first to leave the softly-broken global SUSY theory.

In the last section, we have demonstrated that the fine-tuning of the Higgs
doublet masses in the SUSY-GUT is not preserved if the most general
soft-terms are switched on, and there arises a large mixing mass term
for the scalars.   As will be seen later, our ansatz (\ref{ansatz})
avoids the emergence of the dangerous terms from the heavy sector.  On
the other hand, if we take $U \neq AW$ the gauge hierarchy achieved by
the fine-tuning will be violated by the soft-terms.  We will discuss
this in section \ref{subsec:stability}.

The scalar potential we consider is summarized as follows
\begin{eqnarray}
    V &=& V_{SUSY} +
 V_{    \begin{picture}(25,0)(0,0)
             \put(0,0){\scriptsize $SUSY$}
             \put(0,0){\line(4,1){22}}
             \end{picture} }
\label{Vfull} \\
    V_{    \begin{picture}(25,0)(0,0)
             \put(0,0){\scriptsize $SUSY$}
             \put(0,0){\line(4,1){22}}
             \end{picture} }
     &=& \{ m_S A W(z)
       +m_S F^\kappa B^*_\kappa (z^*) +h.c. \}
\nonumber \\
       & & + m_S^2 C(z,z^*).
\label{Vsoft}
\end{eqnarray}
Note that the scalar potential (\ref{Vfull}) is rewritten as
\begin{eqnarray}
    V &=& - F^*_\kappa F^\kappa
      +\{ F^\kappa \frac{\partial W}{\partial z^\kappa}+h.c. \}
\nonumber \\
 &-& \frac{1}{2} D^\alpha D^\alpha +D^\alpha (z^\dagger T^\alpha z)
\nonumber \\
    &+&\{ m_S A W(z)
     + m_S B^\kappa (z) \frac{\partial W}{\partial z^\kappa}  +h.c. \}
\nonumber \\
       & +& m_S^2 B_\kappa^* (z^*) B^\kappa (z) + m_S^2 C(z,z^*)
\end{eqnarray}
by shifting the auxiliary field $F^\kappa$ as $F^\kappa \rightarrow
F^\kappa +m_S B^\kappa (z)$.  This is the form given at the last section.

\subsection{Calculation of the Effective Potential}

In this subsection we compute the scalar potential of the effective
low-energy
theory by substituting the heavy fields with the solutions to the stationary
conditions of the full potential.  For this aim, it is convenient to
write both the superpotential and the soft SUSY breaking terms in
terms of the variations $\Delta z^{\kappa}=z^\kappa - z_0^\kappa$ in
place of the scalar fields $z^{\kappa}$ themselves.  For the
renormalizable theories, the superpotential can always be written as
\begin{equation}
W = {1 \over 2!}\mu_{\kappa\lambda}\Delta z^{\kappa}\Delta z^{\lambda}
      + {1 \over 3!}h_{\kappa\lambda\mu}\Delta z^{\kappa}
      \Delta z^{\lambda}\Delta z^{\mu}.
	\label{W}
\end{equation}
The functions $B^\kappa(z)$, $B^*_\kappa(z^*)$ and $C(z,z^*)$ are also
expanded as
\begin{eqnarray}
    B^\kappa (z)&=& B_1^\kappa + B_{2 \lambda}^\kappa \Delta z^\lambda
\\
    B^*_\kappa (z^*)&=& B^*_{1 \kappa}
        +B^{* \lambda}_{2 \kappa} \Delta z^*_\lambda
\\
    C(z,z^*)&=& C_{1 \kappa} \Delta z^\kappa
               +C_1^\kappa \Delta z^*_\kappa
\\
            &+& C_{2 \lambda}^\kappa \Delta z^*_\kappa \Delta z^\lambda
               +\frac{1}{2} C_{2 \kappa \lambda}
                \Delta z^\kappa \Delta z^\lambda
               +\frac{1}{2} C_2^{\kappa \lambda}
                \Delta z^*_\kappa \Delta z^*_\lambda .
\end{eqnarray}
 From Eq.~(\ref{Bbilinear}), it follows that $B_1^\kappa=B_{2 \lambda}^\kappa
z_0^\lambda$.

The variations of the potential (\ref{Vfull}) with respect to the
auxiliary fields $F$, $D$ and the scalar fields $z$ are given as
\begin{eqnarray}
\frac{\partial V}{\partial F^\kappa} &=&
	- F_\kappa^\ast + \frac{\partial W}{\partial z^\kappa}
        +m_S B^*_\kappa (z^*)
\nonumber \\
    &=& -F^*_\kappa +\mu_{\kappa \lambda} \Delta z^\lambda
      +\frac{1}{2}h_{\kappa \lambda \mu} \Delta z^\lambda \Delta z^\mu
\nonumber \\
    & & +m_S (B^*_{1 \kappa} + B^{* \lambda}_{2 \kappa} \Delta z^*_\lambda),
\label{F}	\\
\frac{\partial V}{\partial D^\alpha} &=&
	- D^\alpha + z^\dagger T^\alpha z,
		\label{D}\\
\frac{\partial V}{\partial z^\kappa} &=&
	 D^\alpha (z^\dagger T^\alpha)_\kappa
     +\frac{\partial^2 W}{\partial z^\kappa \partial z^\lambda}
      F^\lambda
\nonumber \\
     & & + m_S A \frac{\partial W}{\partial z^\kappa}
         +m_S \frac{\partial B^\lambda}{\partial z^\kappa} F^*_\lambda
         +m_S^2 \frac{\partial C}{\partial z^\kappa}
\nonumber \\
     & =&  D^\alpha (z^\dagger T^\alpha)_\kappa
	+ (\mu_{\kappa\lambda} + h_{\kappa\lambda\mu}\Delta z^{\mu})
        F^{\lambda}
\nonumber\\
     & & +m_S A (\mu_{\kappa \lambda} \Delta z^\lambda
               +h_{\kappa \lambda \mu} \Delta z^\lambda \Delta z^\mu)
\nonumber \\
     & & +m_S B_{2 \kappa}^\lambda F^*_\lambda
         +m_S^2 (C_{1 \kappa} +C_{2 \kappa}^\lambda \Delta z^*_\lambda
                +C_{2 \kappa \lambda} \Delta z^\lambda)
	\label{delV}
\end{eqnarray}

Once the SUSY breaking terms are turned on, the $F^\kappa$ or $D^\alpha$
may be non-vanishing, but should be at most $O(m_{S} M_X)$ since they
have to vanish in the absence of the SUSY breaking terms.
We expand
the $F^\kappa$, $D^\alpha$ and $z^\kappa$ in powers of $m_{S}$
 such as
\begin{eqnarray}
F^\kappa &=& F^\kappa_0 + \delta F^\kappa + \delta^2 F^\kappa +
\cdots,
\label{Fexp}\\
D^\alpha &=& D^\alpha_0 + \delta
D^\alpha + \delta^2 D^\alpha + \cdots,
\label{Dexp}\\
z^\kappa &=&
z^\kappa_0+\Delta z= z^\kappa_0 + \delta z^\kappa + \delta^2 z^\kappa
+ \cdots, 	\label{zexp}
\end{eqnarray}
with $\delta^n F^\kappa, \delta^n D^\alpha = O(m_{S}^n / M_X^{n-2})$
and $\delta^n z^\kappa = O(m_{S}^n / M_X^{n-1})$.
Here $F^\kappa_0$ and $D^\alpha_0$
%and $z^\kappa_0$
are defined as the VEVs in the absence of the SUSY breaking terms and are
exactly zero as discussed in section 2.1.1, and the higher order terms are
defined as the shifts of their VEVs due to
the presence of the SUSY breaking terms.  We assume $z^k =
O(m_{S})$ for the light fields, {\it e.g.},\/ $z^k_0 = O(m_S)$ and
$\delta^2 z^k = \delta^3 z^k = \cdots =0$.

We can solve the stationary conditions (\ref{F})--(\ref{delV})
 by using the above expansions (\ref{Fexp})--(\ref{zexp})
order by order.   Here we list the equations which are to be used to obtain
the scalar  potential of the effective theory.
For simplicity we assume
\begin{equation}
     B_1^k=O(m_S)
\end{equation}
for the time being.  This is  automatically satisfied if there is no
light singlet field.
The stationary conditions $\partial V/\partial F^K=0$ and $\partial
V/\partial F^A=0$ imply
\begin{equation}
   \delta F^*_K=\mu_{KL} \delta z^L+m_S B^*_{1 K}
\label{stFK1}
\end{equation}
and
\begin{eqnarray}
    	\delta F_A^{\ast} &=& m_S B_{1A}^* ,
\label{stFA1}\\
	\delta^2 F_A^{\ast}
     &=& {1 \over 2}h_{A\lambda \mu} \delta z^{\lambda}\delta z^{\mu}
         +m_{S} B_{2A}^\lambda \delta z^*_\lambda
\nonumber \\
      &=& \frac{1}{2}h_{AKL} \delta z^K \delta z^L
         +h_{AKk}\delta z^K \delta z^k
         +m_{S} B_{2A}^\lambda \delta z^*_\lambda
\label{stFA2}
\end{eqnarray}
respectively,\footnote{ Gauge invariance implies
$h_{Akl}=O(m_{S}/M_X)$, see Appendix \ref{app:gauge-inv}.  We have
also used $\delta z^A=0$, which will be derived below.}
while \( \partial V /\partial D^A =0 \) gives
\begin{equation}
	\delta D^A = (z_0^{\dagger}T^A)_B\delta z^B +
                \delta z_B^{\ast}(T^Az_0)^B.
\label{stD1}
\end{equation}
 From the conditions \( \partial V/\partial z^K=0 \) and
\( \partial V /\partial z^A=0 \), we find
\begin{eqnarray}
  &	-\mu_{KL} \delta F^L =& 0
\label{stz1}\\
  &	-\mu_{KL} \delta^2 F^L =& h_{K\lambda\mu}
         \delta F^{\lambda}
	 \delta z^{\mu} + \delta D^A(\delta z^{\dagger} T^A)_K
\nonumber\\
	 & & +m_S A \mu_{KL} \delta z^L
           +m_S B_{2 K}^\lambda \delta F^*_\lambda
           +m_S^2 C_{1K},
\label{stz2}
\end{eqnarray}
and
\begin{eqnarray}
        \delta D^B(z^{\dagger}_0T^B)_A&=&0,
\label{stzA1}\\
	 - \delta^2 D^B(z^{\dagger}_0T^B)_A&=&
   	 h_{A\lambda\mu}\delta F^{\lambda}
    	\delta z^{\mu}
        +m_S B_{2A}^\lambda \delta F^*_\lambda +m_S^2 C_{1A},
\label{stzA2}
\end{eqnarray}
respectively.

{}From Eqs.~(\ref{stz1}) and (\ref{stFK1}), we find the
shift of $z^K$ is
\begin{equation}
   \delta z^K =-m_S (\mu^{-1})^{KL} B^*_{1L}. \label{shiftz}
\end{equation}
On the other hand, Eqs.~(\ref{stD1}) and (\ref{stzA1}) imply
\begin{equation}
\delta z^A=0.
\end{equation}
   Eq.~(\ref{stz2}) gives the solution for $\delta^2
F^K$ as
\begin{eqnarray}
   \delta^2 F^K &=& \langle F^K \rangle
       -m_S (\mu^{-1})^{KL}h_{LAl}B_1^A  \delta z^l,
\\
   \langle F^K \rangle &=& -(\mu^{-1})^{KL} \{ m_S h_{LAM} B_1^A \delta z^M
     +m_S A \mu_{LM} \delta z^M
\nonumber \\
     & & +m_S^2 B_{2L}^A B^*_{1A} +m_S^2 C_{1L} \}.
\end{eqnarray}
 From Eq.~(\ref{stzA2}),
\begin{equation}
	 - \delta^2 D^B(z^{\dagger}_0T^B)_A =
   	m_S h_{A B K} B_1^B
    	\delta z^K
        +m_S^2 B_{2A}^B B_{1B}^* +m_S^2 C_{1A},
\label{Dterm}
\end{equation}
where we have used $h_{ABk}=O(m_S/M_X)$, a consequence of the gauge
invariance.   Note that $(z_0^\dagger T^B)_A = \frac{1}{\sqrt{2}}
(M_V^*)_{BA}$ can be inverted to obtain $\delta^2 D^B$.
Eq.~(\ref{Dterm}) shows that $ \delta^2 D^A$ is a
constant independent of the light fields.  Therefore we will denote it
by $\langle D^A \rangle$.

Now it is straightforward to calculate the scalar potential of the
low-energy effective theory $V_{\it eff}$ by substituting the solutions
to the stationary conditions for the heavy fields.  The result can be
compactly expressed if we define
\begin{eqnarray}
    \widetilde{W} (z) &=&
     W(z^k, z^K=z_0^K+\delta z^K , z^A=z_0^A+ \delta z^A),
\label{effectiveW}\\
    \widetilde{B}^k(z)& =&
         B^k(z^l, z^K=z_0^K+\delta z^K, z^A=z_0^A +\delta z^A),
\\
    \widetilde{C}(z,z^*)& =&
    C(z^k, z^K=z_0^K+\delta z^K , z^A=z_0^A+ \delta z^A,
      \cdots ).
\end{eqnarray}
Note that the above are the functions of only light fields.  In
particular the $\widetilde W$ is the superpotential of the effective
theory.  Then we can write down the effective potential as
\begin{eqnarray}
    V_{\it eff}&=& -F^*_k F^k
        +F^k \left(\frac{\partial \widetilde{W}}{\partial z^k}
             +m_S \widetilde{B}^*_k \right) +h.c.
\nonumber \\
     & & -\frac{1}{2} D^a D^a +D^a (z^\dagger T^a z)
\nonumber \\
     & & +m_S A \widetilde{W}(z) +h.c. +m_S^2 \widetilde{C}(z,z^*)
\nonumber \\
    & & +\Delta V,
\label{Veff}
\end{eqnarray}
where the new contribution $\Delta V$ is
\begin{eqnarray}
    \Delta V &=& -\left| \langle F^K \rangle
                 -m_S (\mu^{-1})^{KL}h_{LAk}B_1^A \delta z^k \right|^2
\nonumber \\
    & & +\left\{ \langle F^K \rangle
     \left(\frac{1}{2}h_{K \lambda \mu} \delta z^\lambda \delta z^\mu
      +m_S B_{2K}^\lambda \delta z^*_\lambda\right) \right.
\nonumber \\
    & & \ \ -m_S (\mu^{-1})^{KL} h_{LAk} B_1^A \delta z^k
        \left(\frac{1}{2}h_{K \lambda \mu} \delta z^\lambda \delta z^\mu
      +m_S B_{2K}^\lambda \delta z^*_\lambda\right)
\nonumber \\
    & & \ \ \left. +h.c. \right\}
\nonumber \\
    & & + \left|m_S B^*_{1A} +m_S B^{*K}_{2A} \delta z^*_K
        + \frac{1}{2} h_{AKL} \delta z^K \delta z^L \right.
\nonumber \\
    & & \ \ \  \ \ \left. + m_S B^{*k}_{2A} \delta z^*_k
        +h_{AKl} \delta z^K \delta z^l \right|^2
\nonumber \\
   & & +\langle D^A \rangle \delta z^\dagger T^A \delta z.
\label{Vnew}
\end{eqnarray}
Recall that $\langle D^A \rangle$ stands for $\delta^2 D^A$ (see
Eq.~(\ref{Dterm})).   The last
term in Eq.~(\ref{Vnew}) comes from the $D$-term of the heavy
gauge sector and is referred to as the $D$-term contribution, while the
other contributions are called the $F$-term contributions.

Eliminating the auxiliary fields $F^k$ and $D^a$ by using the
equations of motion, we obtain the SUSY breaking part of the effective
potential
\begin{eqnarray}
    V_{{\it eff},    \begin{picture}(25,0)(0,0)
             \put(0,0){\scriptsize $SUSY$}
             \put(0,0){\line(4,1){22}}
             \end{picture} }
 &=& m_S  A \widetilde{W} (z)
       +m_S \widetilde{B}^k(z) \frac{\partial \widetilde{W}}{\partial z^k}
+h.c.
 \nonumber \\
& &    +m_S^2 \{\widetilde{B}^*_k(z^*) \widetilde{B}^k(z)
     +\widetilde{C}(z,z^*) \} +\Delta V.
\label{Veff:breaking}
\end{eqnarray}

\subsection{Stability of the Weak Scale}
\label{subsec:stability}
In this subsection, we will investigate whether the effective
potential is of the order of magnitude $m_S^4$.\footnote{ Here we
disregard the constant terms independent of the light fields.} The
occurrence of the terms of $O(m_S^3 M_X)$ or even larger is very
dangerous since it would destabilize the weak scale.  Such dangerous
terms may appear for a mixing mass term (proportional to $\delta
z^k \delta z^l$) and for a linear term (proportional to $\delta z^k$).
The latter  is related to the
notorious difficulty in the presence of the light singlet \cite{PS}. Note
that the linear term cannot exist for a non-singlet field.   Whether
the linear term is actually large or not is highly model dependent.  We
will not discuss this problem further.   In the rest of this section we will
concentrate on the mixing mass term.

{}From Eqs.~(\ref{Veff:breaking}) and (\ref{Vnew}), it follows
that the mixing mass terms are
\begin{eqnarray}
& & m_S A \mu_{k l} +m_S h_{klm} (B_1^m +B_{2M}^m \delta z^M)
\nonumber \\
& & +m_S (B_{2k}^m \mu_{ml} +B_{2l}^m \mu_{mk})
+ m_S^2 C_{2kl}+\langle F^K \rangle h_{Kkl}
\nonumber \\
& & -m_S(\mu^{-1})^{KL} h_{KAk} B_1^A h_{LMl} \delta z^M
-m_S(\mu^{-1})^{KL} h_{KAl} B_1^A h_{LMk} \delta z^M
\nonumber \\
& & + m_{S} h_{KAk} \delta z^K B_{2l}^{*A}
    + m_S h_{KAl} \delta z^K B_{2k}^{*A},
\end{eqnarray}
which are of order $m_S^2$.
 This is due to our
assumptions (1) $U=AW$ and (2) $B^k_1=O(m_S)$.  We will show that
larger terms arise when we relax the assumptions.

First consider the case of $U \neq AW$.  An inspection similar to the
previous subsection shows that there exist terms of order $m_S M_X$
for the mixing mass
\begin{equation}
   \langle \frac{\partial^2 U}{\partial z^k \partial z^l} \rangle
   +\langle F^K \rangle h_{Kkl}+O(m_S^2), \label{large-mixing}
\end{equation}
where
\begin{equation}
     \langle F^K \rangle =-m_S (\mu^{-1})^{KL}
     \langle \frac{\partial U}{\partial z^L} \rangle +O(m_S^2).
\end{equation}
The first term in Eq.~(\ref{large-mixing}) can be $O(m_S M_X)$,
 since there is no {\it a priori} reason that the fine-tuning of the
Higgs mass achieves simultaneously both in $W$ and $U$.  The second
term remains large  if the conditions $\partial W/\partial z^K=0$ and
$\partial U/\partial z^K=0$ do not hold simultaneously.  This
observation shows the importance of our ansatz $U=AW$.

Next consider the effects of $B_1^m$ of order $M_X$.  Obviously
$B_1^m$ is non-zero only for a (light) singlet.  Then we find an
additional contribution to the mixing mass term
\begin{equation}
     m_S B_1^m h_{klm}.
\end{equation}
 This again reflects the well-known difficulty related to the light
singlet.  Indeed a large $B$ term can appear at the tree-level or may be
induced by radiative corrections\cite{PS,BP}.  To proceed further we need a
model
dependent analysis which is beyond the scope of the
paper.\footnote{For example, in a flipped $SU(5)$ model \cite{AEHN}, it
is known that the light singlet which couples to the Higgs doublets does not
induce the large mixing mass.}

To conclude, the large mixing mass terms do not arise if (1) one takes
the hidden assumption and thus $U=AW$, and (2) there is no light singlet.

\subsection{Mass Terms}
\label{subsec:mass-terms}
We now discuss a {\em chirality-conserving} mass term, namely the
coefficient of $\delta z^k \delta z^*_l$.  They are easily extracted
{}from  Eqs.~(\ref{Veff}) and (\ref{Vnew}), given by
\begin{eqnarray}
  & &  m_S^2 (B_{2k}^{m} B_{2m}^{*l}+C_{2k}^l)
  -m_S^2 (\mu^{-1})^{MK} h_{KAk} B_{1}^A
     (\mu^{* -1})_{ML} h^{*LBl} B_{1B}^*
\nonumber \\
  & & -m_S^2 (\mu^{-1})^{KL} h_{KAk} B_1^A B_{2L}^l
     -m_S^2 (\mu^{*-1})_{KL} h^{*LAl} B_{1B}^* B_{2k}^{*K}
\nonumber \\
   & & +m_S^2 B_{2k}^A B^{*l}_A
       +h_{AKk} h^{*ALl} \delta z^K \delta z^*_L
\nonumber \\
   & & +\langle D^A \rangle (T^A)^l_k. \label{chirality-conserving-mass}
\end{eqnarray}
The term $m_S^2 (B_{2k}^m B^{*l}_{2m}+C_{2k}^l)$ is present before the
heavy sector is integrated out.  Therefore it respects the large gauge
symmetry of the unified group.  On the other hand, other terms coming
 from $\Delta V$ can pick up effects of the symmetry breaking.

The last term in Eq.~(\ref{chirality-conserving-mass}) is the $D$-term
contribution.  Phenomenologically it is important because it gives
an additional contribution to squarks and sleptons \cite{KMY,KT}.  Gauge
invariance implies that the non-zero VEV of the $D$-term is allowed for
the $U(1)$ factor.  Thus it can arise when the rank of the gauge group
is reduced by the gauge symmetry breaking.  The $D$-term contribution is
proportional to the charge of the broken $U(1)$ factor and gives
mass splittings within the same multiplet in the full theory.

We can rewrite $\delta^2 D^A=\langle D^A \rangle $ by
using the gauge invariance of $W$ and $B$ as
\begin{eqnarray}
  \langle D^A \rangle & =& - 2 m_S^2 (M_V^2)^{-1 AB}
       \{B_{1 \kappa}^* (T^B)^\kappa_\lambda  B_1^\lambda
        -B_{1 K}^* (T^B)^K_L B_{1}^L
\nonumber \\
      & & + C_{2 \kappa}^\lambda z_{0 \lambda}^* (T^B z_0)^\kappa
          +C_{2 \kappa \lambda} z_0^\lambda (T^B z_0)^\kappa \}.
\label{VEVD}
\end{eqnarray}
We will see in the next subsection that the VEV of the $D$-term
(\ref{VEVD}) vanishes when the SUSY breaking terms are universal
and hence the $D$-term contribution to the sfermion masses is
a characteristic of  the non-universal SUSY breaking.

\subsection{Case of the Universal Soft SUSY Breaking Terms}
Let us now discuss the effective potential for the case of the
universal soft terms.  In addition to our ansatz
(\ref{ansatz}),  we assume that
\begin{eqnarray}
     B(\Phi, \Phi^\dagger)&=& B\Phi^\dagger \Phi,
\\
     C(\Phi, \Phi^\dagger)&=& C \Phi^\dagger \Phi,
\end{eqnarray}
with dimensionless constants $B$ and $C$.   Then we find
\begin{equation}
    B_1^\kappa =B z^\kappa_0, \ \ \
    B_{2\lambda}^\kappa=B \delta^\kappa_\lambda
\end{equation}
and
\begin{equation}
    C_{1 \kappa}=C z_{0 \kappa}^*, \ \  C_1^\kappa=C z^\kappa_0, \ \
    C_{2 \lambda}^\kappa= C \delta ^\kappa_\lambda, \ \
    C_{2 \kappa \lambda}=C_2^{\kappa \lambda}=0.
\end{equation}
In particular,  $B_1^A=C_1^A=0$ because $z_0^A=0$.  Furthermore our
assumption $B_1^k =O(m_S)$ is automatically satisfied since we
postulate $z_0^k=O(m_S)$, which allows us to use the effective
potential (\ref{Veff}) and (\ref{Vnew}).

We first evaluate $\Delta V$ in Eq.~(\ref{Vnew}).
Eq.~(\ref{shiftz}) now reads
\begin{equation}
   \delta z^K=-m_S (\mu^{-1})^{KL} B z_{0L}^*.
\end{equation}
The VEVs of the auxiliary fields become
\begin{eqnarray}
   \langle F^K \rangle
    &=& -m_S A \delta z^K -m_S^2 (\mu^{-1})^{KL} C z^*_{0L},
\\
    \langle D^A \rangle
   &=& 2 m_S^2 (M_V^2)^{-1 AB} C (z_0^\dagger T^A z_0)=0,
\end{eqnarray}
where the last equality is due to Eq.~(\ref{SUSYD}).  Noting that the gauge
invariance of the superpotential shows
\begin{eqnarray}
     h_{BKl} \delta z^K (T^A z_0)^B
   &=& -\mu_{KL} \delta z^L (T^A)^K_l +O(m_S^2)
\nonumber \\
   &=& m_S B (z_0^\dagger T^A)_l +O(m_S^2)
\nonumber \\
   &=& 0,
\end{eqnarray}
we find $\Delta V$ becomes simply
\begin{equation}
   \Delta V= \frac{1}{2} \langle F^K \rangle h_{K \lambda \mu}
                \delta z^\lambda \delta z^{\mu}.
\end{equation}
Hence all contributions to the {\em chirality-conserving} mass terms
disappear with the universal soft terms.  The SUSY breaking part of
the potential is found to be
\begin{eqnarray}
 V_{{\it eff},    \begin{picture}(25,0)(0,0)
             \put(0,0){\scriptsize $SUSY$}
             \put(0,0){\line(4,1){22}}
             \end{picture} }
&=& m_S A \widetilde{W} +m_S B z^k
         \frac{\partial \widetilde{W}}{\partial z^k}
          +\frac{1}{2} \langle F^K \rangle h_{K\lambda \mu}
           \delta z^\lambda \delta z^\mu +h.c.
\nonumber \\
  & &   +m_S^2 (B^2+C)z^*_k z^k.
\end{eqnarray}
An important conclusion is that the scalar mass is common in this
case.

To compare our results with those of Ref. \cite{HLW},  we further take
\begin{equation}
    B=1, \ \ \ C=0,
\end{equation}
as well as $\mu_{kl}=0$, $z^k_0=0$.  Then after a little algebra, we find
\begin{eqnarray}
   \Delta V &=&
   -\frac{1}{2} m_S A h_{K \lambda \mu} \delta z^K \delta z^\lambda
     \delta z^\mu
\nonumber \\
  & =&  -m_S A \left(3 \widetilde{W} -z^k \frac{\partial \widetilde W}{\partial
     z^k}\right),
\end{eqnarray}
and the SUSY breaking part is
\begin{equation}
 V_{{\it eff},    \begin{picture}(25,0)(0,0)
             \put(0,0){\scriptsize $SUSY$}
             \put(0,0){\line(4,1){22}}
             \end{picture} }
=-2 m_S A \widetilde{W} +m_S (A+1) z^k
         \frac{\partial \widetilde{W}}{\partial z^k}
        +h.c.  +m_S^2 z^*_k z^k,
\end{equation}
which is in agreement with the result of Ref. \cite{HLW}.

\section{Phenomenological Implications}
\cleqn
In this section, we point out phenomenological implications
of the general discussion in the previous sections. The main new feature is
that one may have non-universal scalar masses at the GUT scale, both
due to the $D$-term and $F$-term contributions. For first two
generation slepton/squark fields, we expect that the superpotential
coupling is weak enough to be neglected, and we deal with only the
$D$-term contributions. It was pointed out in our previous
paper \cite{KMY,KT} that the scalar masses satisfy ``sum-rules''
corresponding to the symmetry breaking pattern of the grand unified
theory, which can be tested at future collider experiments. On the
other hand, the Higgs fields and third generation slepton/squark
fields are likely to acquire both $D$- and $F$-term contributions,
which may drastically change the analysis of the radiative breaking
scenario.

\subsection{Squark and Slepton Masses}
\label{subsec:sfermion}

For the first two generation matter fields, the superpotential coupling
is small and can be neglected. Then the masses of their scalar
components are determined solely by the initial conditions and their
gauge quantum numbers. As a consequence, they have a definite pattern in
the mass spectrum once one has a specific symmetry breaking pattern from
the grand-unified group down to the Standard Model gauge group, $G_{SM}
= SU(3)_C \times SU(2)_L \times U(1)_Y$.

In our previous paper \cite{KMY}, we showed that the squark and slepton masses
satisfy certain ``sum rules'' for various examples of the symmetry
breaking patterns. Let us briefly review the results below, as an
example where the $D$-term contributions to the scalar masses play a major
phenomenological role.

Let us take the following symmetry breaking pattern for instance:
\newcommand{\verylongrightarrow}{
\relbar\joinrel\relbar\joinrel\relbar\joinrel\rightarrow}
\newcommand{\breaksto}[1]{\mathop{\verylongrightarrow}\limits^{#1}}
\begin{equation}
SO(10) \breaksto{M_{U}} SU(4)_{PS} \times SU(2)_L \times SU(2)_R
	\breaksto{M_{PS}} G_{SM}.
\end{equation}
There appear $D$-term contributions to the scalar masses when the rank
of the gauge group reduces from 5 to 4 at the intermediate Pati-Salam
symmetry breaking scale $M_{PS}$. The matter multiplets belong either to
$L=({\bf 4}, {\bf 2}, {\bf 1})$ or $R=({\bf \bar 4}, {\bf 1}, {\bf 2})$
representations, with masses $m_L^2$ and $m_R^2$ respectively above
$M_{PS}$. When the Pati-Salam group breaks to $G_{SM}$, we obtain the
following masses,
\begin{eqnarray}
m_{\tilde{q}}^2 &=& m_L^2 + g_4^2 D,\\
m_{\tilde{u}}^2 &=& m_R^2 - (g_4^2 - 2 g_{2R}^2) D,\\
m_{\tilde{e}}^2 &=& m_R^2 + (3 g_4^2 - 2 g_{2R}^2) D,\\
m_{\tilde{l}}^2 &=& m_L^2 - 3 g_4^2 D,\\
m_{\tilde{d}}^2 &=& m_R^2 - (g_4^2 + 2 g_{2R}^2) D,
\end{eqnarray}
where $D$ represents the $D$-term contributions whose normalization is
taken arbitrarily. These expressions do not depend on a particular choice
of the Higgs representation which breaks Pati-Salam group, and hence
fixed completely by the symmetry breaking pattern.\footnote{One may
anticipate that there are also $F$-term contributions since we need a
superpotential $RR\chi$ to generate right-handed neutrino masses where
$\chi$ transforms as $({\bf \bar{15}}, {\bf 1}, {\bf 3})$. Indeed,
right-handed scalar neutrino may acquire contributions like $F_\chi
\tilde{N}^c \tilde{N}^c$. However other squark, slepton fields cannot
acquire $F$-term contribution because of the $G_{SM}$ invariance.} Note that
the gauge coupling constants $g_4^2$, $g_{2R}^2$ can be determined from
the low-energy gauge coupling constants $\alpha_i (m_Z)$ $(i=1,2,3)$ as
a function of $M_{PS}$ alone. On the other hand, one can eliminate $D$,
$m_L^2$ and $m_R^2$ form the above formulae, to obtain
\begin{eqnarray}
m_{\tilde{q}}^2 (M_{PS}) - m_{\tilde{l}}^2 (M_{PS})
        &=& m_{\tilde{e}}^2 (M_{PS}) - m_{\tilde{d}}^2 (M_{PS}),
                \label{PS1} \\
g_{2R}^2 (M_{PS}) (m_{\tilde{q}}^2 - m_{\tilde{l}}^2) (M_{PS})
        &=& g_4^2 (M_{PS}) (m_{\tilde{u}}^2 - m_{\tilde{d}}^2) (M_{PS}).
                \label{PS2}
\end{eqnarray}
Once we measure the gaugino and scalar masses at low-energy, we can
calculate the scalar masses at $M_{PS}$ as a function of $M_{PS}$ alone.
Since we have two relations for one free parameter $M_{PS}$, one can
solve for $M_{PS}$ and further make a consistency check.

There exist more relations when $SO(10)$ is broken directly into
$G_{SM}$,
\begin{eqnarray}
m_{\tilde{q}}^2 &=& m_{16}^2 + g_{10}^2 D,\\
m_{\tilde{u}}^2 &=& m_{16}^2 + g_{10}^2 D,\\
m_{\tilde{e}}^2 &=& m_{16}^2 + g_{10}^2 D,\\
m_{\tilde{l}}^2 &=& m_{16}^2 - 3 g_{10}^2 D,\\
m_{\tilde{d}}^2 &=& m_{16}^2 - 3 g_{10}^2 D,
\end{eqnarray}
where the only unknown parameters are $m_{16}^2$ and  $D$ after the
measurements of the SUSY breaking
masses. We use one relation to fix $D$, one for $m_{16}^2$, and
there remain three relations for the
consistency check. On the other hand, we have more free parameters when
the symmetry is smaller, for instance,
\begin{equation}
SO(10) \breaksto{M_U} SU(3)_C \times SU(2)_L \times SU(2)_R \times
U(1)_{B-L} \breaksto{M_{3221}} G_{SM}.
\end{equation}
In this case we cannot even determine the parameters in the original
model.

On the other hand, let us consider flipped $SU(5)$ model \cite{flip,AEHN}
as an example of non-unified model.  Its gauge group is $SU(5) \times
U(1)$ and is not grand-unified into a single group.  First of all, we
have two independent gaugino masses $M_{SU(5)}$ and $M_{U(1)}$ at the
GUT scale, which results in the low-energy gaugino masses which do not
necessarily satisfy the GUT-relation.  However, one can test the scenario
measuring $M_2$ and $M_3$ at the weak scale and see whether they unify
at the same scale where the gauge coupling constants $\alpha_2$ and
$\alpha_3$ unify. On the scalar masses, we have three independent
masses $m_{10}^2$, $m_{\bar 5}^2$ and $m_{1}^2$ at the GUT scale, and
an unknown $D$-term in addition.  There are five observable scalar
masses, and we know the scale where $SU(2)$ and $SU(3)$ coupling
constants meet. Therefore we are left with one additional relation
which can be checked. The scalar masses satisfy the following
relations at $M_X$ where $SU(2)$ and $SU(3)$ unify to $SU(5)$,
\begin{eqnarray}
{m}_{\tilde{d}}(M_{X})^2 &=& {{m}_{10}}^2 +
\left(-{2 \over 5}{g_{SU(5)}}^2 + {1 \over 40}{g_{U(1)}}^2 \right) D, \\
{m}_{\tilde{l}}(M_{X})^2 &=& {{m}_{\bar{5}}}^2 -
\left({3 \over 10}{g_{SU(5)}}^2 + {3 \over 40}{g_{U(1)}}^2 \right) D, \\
{m}_{\tilde{u}}(M_{X})^2 &=& {{m}_{\bar{5}}}^2 +
\left({1 \over 5}{g_{SU(5)}}^2 - {3 \over 40}{g_{U(1)}}^2 \right) D, \\
{m}_{\tilde{q}}(M_{X})^2 &=& {{m}_{10}}^2 +
\left({1 \over 10}{g_{SU(5)}}^2 + {1 \over 40}{g_{U(1)}}^2\right) D, \\
{m}_{\tilde{e}}(M_{X})^2 &=& {{m}_{1}}^2 +
{1 \over 8}{g_{U(1)}}^2 D .
\end{eqnarray}
Then the ``sum rule'' is obtained as
\begin{equation}
m_{\tilde{d}}^2 (M_X) - m_{\tilde{l}}^2 (M_X)
	= m_{\tilde{q}}^2 (M_X) - m_{\tilde{u}}^2 (M_X).
\end{equation}

Though the ``sum rules'' of the scalar masses are weak when the symmetry
is small, one can acquire useful information by combining the scalar
mass spectrum with that of the gauginos. We pointed out that the gaugino
masses satisfy the so-called GUT-relation
\begin{equation}
\frac{M_1}{\alpha_1} = \frac{M_2}{\alpha_2} = \frac{M_3}{\alpha_3},
\end{equation}
even when the grand-unified group breaks down to $G_{SM}$ in several
steps. On the other hand, non-GUT models like flipped $SU(5)$ do not
necessarily have a unified gaugino mass, and hence do not predict
GUT-relation of the gaugino masses. Therefore, one can draw useful
information on the GUT models by measuring the scalar and gaugino masses
in future experiments. We present a ``score sheet'' of various models in
Table 1. Here we refer to the paper \cite{BIM} on superstring
predictions.\footnote{It is noteworthy that the superstring with
dilaton $F$-term also leads to the same relation. This is
amusing because one needs rather big threshold corrections for the gauge
coupling constants to reconcile the difference between the apparent
GUT-scale and the string scale \cite{ILR}. Exactly the same correction
appears both for the gauge coupling constants and the gaugino masses to
give the same relation as in the (field theoretical) GUT models.}

%\begin{table}
%\footnotesize
%\begin{tabular}{lccc}
%&$\alpha_i$ & $M_i/\alpha_i$ & $m^2_i$\\ \hline \\
%$SU(5) \rightarrow G_{SM}$ & natural & common & testable\\
%$SO(10) \rightarrow G_{SM}$ & natural & common & testable\\
%$SO(10) \rightarrow G_{PS} \rightarrow
%G_{SM}$ & adjustable & common & testable\\
%$SO(10) \rightarrow G_{3221} \rightarrow G_{SM}$ & adjustable & common &
%not testable\\
%$SO(10) \rightarrow G_{3211} \rightarrow G_{SM}$ & adjustable & common &
%not testable\\
%$SU(5) \times U(1) \rightarrow G_{SM}$ & adjustable & common only for
%$i=2,3$ & testable\\
%superstring with dilaton $F$-term & adjustable & common & testable\\
%superstring with moduli $F$-term & adjustable & not common & not testable
%\end{tabular}
%\caption[1]{The ``score sheet'' how well we can distinguish between
%various models. The intermediate groups are defined as $G_{PS} =
%SU(4)_{PS} \times SU(2)_L \times SU(2)_R$, $G_{3221} = SU(3)_C \times
%SU(2)_L \times SU(2)_R \times U(1)_{B-L}$, $G_{3211} = SU(3)_C \times
%SU(2)_L \times U(1)_R \times U(1)_{B-L}$. The row $\alpha_i$ refers to
%the unification of the gauge coupling constants, where ``natural'' means
%that the unification is automatic, while ``adjustable'' employs either
%particular particle content or threshold corrections to reproduce the
%observed gauge coupling constants. The row $M_i/\alpha_i$ refers to the
%gaugino masses. The row $m^2_i$ states whether the model predicts
%a definite pattern which is testable using the low-energy scalar mass
%spectrum. }
%\end{table}
%

%{\it Upper Bound on the Slepton Mass in No-scale Model}

\subsection{Radiative Breaking Scenario}
\label{subsec:radiative}

There has been reported several remarkable results based on the radiative
breaking scenario \cite{radiative-breaking} and the universal scalar mass
hypothesis at the
GUT scale. One
of them is that the LSP should be gaugino dominant to correctly
reproduce the weak scale \cite{DN}, when top quark is heavy. This gives
a stringent constraint on
the cosmic abundance of the neutralino. The other is that the proton
decay in the minimal $SU(5)$ SUSY-GUT cannot be consistent with the
present bound if one further requires that the LSP does not overclose
the Universe \cite{LNZ}. Since both analyses crucially depend on the universal
scalar mass hypothesis, there may be qualitatively different consequences
in the non-minimal case.

The crucial equations to determine $m_Z$ and $\tan \beta$ in the
minimal SUSY standard model (MSSM) are the
following ones at the tree-level,
\begin{eqnarray}
m_Z^2 &=& - \frac{m_1^2 - m_2^2}{\cos 2\beta} - (m_1^2 + m_2^2 + 2 \mu^2),
	\label{mZ}\\
\sin 2\beta &=& -\frac{2 m_3^2}{m_1^2 + m_2^2 + 2 \mu^2}, \label{beta}
\end{eqnarray}
where $m_1^2$, $m_2^2$ refer to the soft SUSY breaking part of the Higgs
masses, $m_3^2$ the off-diagonal mass, and $\mu$ the higgsino mass
parameter. When one adopts the universal scalar mass hypothesis, the
large top quark Yukawa coupling drives $m_2^2$ much smaller than $m_1^2$
in general as far as $\tan \beta$ is not very large. This gives a too
large
value of the first term in r.h.s of Eq.~(\ref{mZ}) in general, which should be
compensated by the negative contributions of $\mu^2$. Of course the
details depend on the renormalization group analysis, it can be studied
semi-analytically as far as one can neglect the bottom quark Yukawa
coupling constant \cite{ILM}. In this case,
\begin{eqnarray}
m_1^2 &=& m_{\tilde{l}}^2, \\
m_2^2 &=& m_{\tilde{l}}^2 - 3 I,\\
m_3^2 &=& B \mu,
\end{eqnarray}
Here, $I$ is defined by the following,
\begin{equation}
I = I_{SS} m_\infty^2 + I_{GG} M_\infty^2 + I_{GA} M_\infty A_\infty
	+ I_{AA} A_\infty^2,
\end{equation}
with the coefficients $I_{SS}$, $I_{GG}$, $I_{GA}$, $I_{AA}$ are
functions of the top quark Yukawa coupling only, and $m_\infty$, $M_\infty$,
$A_\infty$ are the universal scalar mass, universal gaugino mass, and
universal trilinear coupling at the GUT scale, respectively. One can
rewrite the Eqs.~(\ref{mZ}) as
\begin{eqnarray}
\mu^2 &=& 3 \frac{\tan^2 \beta}{\tan^2 \beta - 1}
	( I_{SS} m_\infty^2 + I_{GG} M_\infty^2 + I_{GA} M_\infty A_\infty
		+ I_{AA} A_\infty^2) \nonumber \\
& &	- (m_\infty^2 + 0.52 M_\infty^2) - m_Z^2/2.
\end{eqnarray}
The coefficient $I_{GG}$ varies from 0.6 to 1.2, and hence $\mu^2$ is
always an increasing function of $M_\infty$. For small $\tan\beta$, the
first term dominates, and one has a large $\mu^2$. For moderately large $\tan
\beta$, the LEP bound on $M_2$ gives a lower bound on $M_\infty$. Though
$m_\infty^2$ and $M_\infty A_\infty$ terms may give negative
contributions to the above equation, their coefficients are in general
not large, and one has to take $m_\infty$ or $|A_\infty|$ very large to
make higgsino light. Since such parameters are not favored from the
naturalness point of view, one reaches the conclusion that the
higgsino-like LSP is disfavored in the radiative breaking scenario. This is
especially true in the no-scale case, where all $m_\infty^2$, $M_\infty
A_\infty$, and $A_\infty^2$ term vanish. It is not possible to obtain
a higgsino-like LSP within the no-scale models. Even with non-vanishing
$m_\infty$ and $A_\infty$, it was shown that there are no solutions with
a higgsino-like LSP \cite{DN} after including the one-loop effects on the
Higgs potential.

However, the situation drastically changes when one incorporates the
possible $D$-term contributions to $m_1^2$ and $m_2^2$. Let us imagine
the initial condition $m_1^2 = m_\infty^2 - \Delta m^2$, $m_2^2 =
m_\infty^2 + \Delta m^2$. This gives an extra contribution $\Delta
\mu^2$ to the $\mu^2$ as
\begin{equation}
\Delta \mu^2 =  \frac{1}{2 \cos 2 \beta}
	(2 - (1-\cos 2\beta) I_{SS}) \Delta m^2,
\end{equation}
which allows a lighter higgsino compared to the case of the universal
scalar mass.

The fact that the lighter higgsino is allowed has a very strong impact
in the minimal $SU(5)$ GUT, where there has been claimed that the
nucleon decay via the dimension-five operators cannot be consistent
with the longevity of the Universe \cite{LNZ}. The nucleon decay rate
is roughly proportional to $M_\infty/m_\infty^2$ when $M_\infty \lsim
m_\infty$ \cite{dimen5}, while gaugino-like LSP has an abundance
proportional to $m_\infty^2/M_\infty$.
However, the abundance becomes much smaller
once the smaller higgsino mass parameter is allowed. Therefore, one
can take $M_\infty$ to be much smaller than $m_\infty$ without
worrying about the LSP abundance, which opens a consistent region
between the nucleon decay experiments and the LSP abundance.

\subsection{FCNC}

The assumption of the universal scalar mass is motivated to explain
the smallness of the flavor-changing neutral current (FCNC) due to the
loops of SUSY particles \cite{FCNC}. Since we have relaxed this
assumption, the readers may be worried about FCNC.

There are two classes of non-minimal effects, one which does not break
the degeneracy between sfermion masses with the same quantum numbers,
the other which does. The non-minimality we discussed in
subsection~\ref{subsec:sfermion} could have been generated by the
renormalization between the Planck scale and the GUT scale. The
implicit assumption here is that the Yukawa interactions are small for
first two generations, even to the superheavy fields which are
completely decoupled from the low-energy effective action. As far as
{\it all}\/ of their interactions in the superpotential are small, the
only renormalization effects arise due to the gauge interactions, and
hence universal for different generations. The degeneracy of sfermion
masses at the initial condition ensures the degeneracy at the weak
scale. In this case, non-minimal nature of the radiative corrections
does not break the degeneracy.

On the other hand, there are many source of radiative corrections which
could break the degeneracy. For instance, there are attempts to explain
the degeneracy of sfermion masses based on horizontal symmetries
\cite{Dine}. If the horizontal symmetry is gauged and breaks
spontaneously, however, there may be $D$-terms in the horizontal gauge
group which potentially breaks the degeneracy again. Another example is
when the first two generations also have $O(1)$ Yukawa interactions
beyond the GUT scale. In this case there are two sources of
non-degeneracy: (1) renormalization due to the Yukawa interactions, (2)
$F$-term contributions to the scalar masses when the heavy particles
decouple. There are no discussions on these effects in the literature to
our knowledge.

It is noteworthy, however, that FCNC can be sufficiently suppressed even
with non-degenerate initial condition. If the scalar masses turn out to
be relatively smaller than the gaugino masses, radiative corrections
between the Planck or GUT scale  and the weak scale tend to make squark masses
universal due to the gluino contribution. Indeed, SUSY breaking via
moduli $F$-term condensation in superstring inspired supergravity models
give non-universal scalar masses which depend on modular weights, and in
principle can lead to large FCNC processes. However, they can be
sufficiently suppressed due to the renormalization effects at least for
some region of the parameter space \cite{BIM}.

\subsection{Renormalization Between GUT and Planck Scales}

As repeatedly emphasized through the text, one of the important sources
of the non-minimality is the renormalization between the GUT and the
Planck scales.
Let us briefly comment when and how these effects can be important
despite the apparent closeness of the two scales. Indeed, most of the
analyses in the literature completely ignore the difference of these two
scales.

One example is when the constraint is marginal. For
instance, it was shown in Refs.~\cite{IKYY,LNY} that one has an upper
bound on the gaugino mass in a restricted class of minimal supergravity
model with $m_\infty = A_\infty = 0$ at the GUT scale. The argument
comes from the fact that the right-handed slepton acquires a mass only of
\begin{equation}
m_{\tilde{l}_R}^2 \simeq 0.87 M_1^2,
\end{equation}
and hence is smaller than $M_1$. Then there is a danger that $\tilde{l}_R$
becomes lighter than the lightest neutralino. One is forced to use
mixing in the neutralino sector to push the LSP mass down from $M_1$.
(Higgsino LSP does not exist as a solution in the radiative breaking
scenario with the universal scalar masses.) However mixing can be substantial
only when gaugino mass is close to $m_Z$, and one obtains an upper bound on
the gaugino mass. It was translated to an upper bound on the slepton
mass, $\lsim 150$~GeV \cite{IKYY,LNY}.
The situation completely changes when one includes the
running of the slepton mass between Planck and GUT scales. Then one obtains
\begin{equation}
m_{\tilde{l}_R}^2 \simeq 3.1 M_1^2,
\end{equation}
and one does not need a substantial mixing in the neutralino sector any
more. Therefore the upper bound on the gaugino mass becomes obsolete due
to this effect.

Another example is when there is a relatively large coupling
constant. If we require $m_b$-$m_\tau$ mass relation, one generally
needs top quark Yukawa coupling constant of $\simeq 2.0$ at the
GUT scale for small $\tan \beta$, and $\simeq 0.8$ for large $\tan \beta
\simeq 60$. Let us take minimal $SU(5)$ for clarify our
discussions. Then $H_u$ has much smaller $m^2$ at the GUT scale compared
to $H_d$ even when they start from the same value at the Planck scale.
One has to go through the following analyses. First one solves the
renormalization group running of the Higgs masses for {\bf 24}, {\bf 5},
${\bf \bar 5}$, and also other SUSY breaking parameters. Then employ the
formulae presented in section 3 to integrate out the superheavy fields to
obtain the low-energy effective action. Such a non-minimality may change
the parameter space of the radiative breaking scenario substantially,
however does not affect FCNC constraint since the effects of $h_t$ does
not appear in the first- and second-generation scalar masses.

%\documentstyle[12pt]{article}
%\begin{document}
\section{Conclusions}
In this paper, we have derived the low-energy effective Lagrangian
 in the scalar sector starting from a unified theory with
 non-universal soft SUSY breaking terms. Such non-universal soft terms
arise if we take a flat limit of the supergravity where the K\"{a}hler
potential is a non-minimal one.   One should note that this is
indeed  the case in the string-inspired model where the moduli fields
are responsible for the SUSY breaking.  Even if the soft
terms have the universal structure at the gravitational scale, they
get renormalized and as a result become non-universal in
general when
the energy scale goes down to the GUT scale.  Therefore we expect that
the soft terms are non-universal at the GUT scale and it is important
to investigate its  consequences at low energies.

We have calculated the scalar potential of the low-energy theory by
explicitly integrating out the heavy sector.  The SUSY breaking part
of the scalar potential is summarized in Eqs.~(\ref{Veff:breaking})
and (\ref{Vnew}).  We found some new contributions to the soft terms
which can be non-zero only when the soft terms of the full theory are
non-universal.  In particular, the sizable $D$-term contributions
generally exist in the chirality conserving scalar masses when the
rank of the gauge group is reduced by the gauge symmetry breaking.  Its
phenomenological implications were discussed in our previous papers
\cite{KMY,KT}.  Another important point is concerned with the gauge
hierarchy problem.  Many of the SUSY GUT models achieve the small
Higgs doublet mass by a fine-tuning of the parameters in the
superpotential.  If the soft terms are turned on, however, a SUSY
breaking Higgs mass term can become heavy and the weak scale will be
destabilized.  We showed that all mass terms remain at the weak scale
if the soft terms are restricted to
those derived from the supergravity model where the hidden sector
decouples from the observable sector in the superpotential.

We have also discussed other phenomenological implications.  Recall
that there are many sources which give the non-universal scalar
masses, including the $D$-term and/or $F$-term contributions discussed
in this paper.  This non-universality changes the predictions of the
radiative electroweak symmetry breaking, usually assuming the common
mass for the two Higgs doublet bosons.  In particular, the higgsino
can be the dominant component of the LSP even when the top quark is
heavy: if we assume the universal scalar mass the LSP is dominated by
a gaugino component.  This cures the apparent conflict of the nucleon
life time and the LSP relic abundance.  In the no-scale model, on the
other hand, we pointed out that the upper bound on the right-handed slepton
mass \cite{IKYY,LNY} disappear if we properly incorporate the
renormalization group flow between the GUT scale and the Planck scale.
Further study of the radiative breaking scenario without the universal
scalar mass hypothesis should be encouraged.

%\end{document}

\cleqn

\section*{Acknowledgements}
We would like to thank T.~Yanagida, K.~Inoue, Y.~Okada and  I.~Jouichi
for useful discussions.  H.M. is also grateful to A.~Brignole,
L.J.~Hall, S.~Dimopoulos, H.~Haber for discussions.
This work was supported in part by the Director, Office of Energy
Research, Office of High Energy and Nuclear Physics, Division of High
Energy Physics of the U.S. Department of Energy under Contract
DE-AC03-76SF00098.

\appendix

\section{Equalities from gauge invariance}
\label{app:gauge-inv}
We summarize the consequence of the gauge invariance of the
superpotential $W$.

 From the gauge invariance of $W$,
there is the equality
\begin{equation}
\frac{\partial W}{\partial z^\kappa} (T^{\alpha})^\kappa_\lambda z^\lambda = 0,
			\label{ginv1}
\end{equation}
where summation over repeated indices $\kappa$ and $\lambda$ is
understood.
If we differentiate the above equality with respect to
$z^\mu$, we obtain
\begin{equation}
\frac{\partial^2 W}{\partial z^\kappa \partial z^\mu}
	(T^{\alpha})^\kappa_\lambda z^\lambda
	+ \frac{\partial W}{\partial z^\kappa}
	(T^{\alpha})^\kappa_\mu = 0.
			\label{ginv2}
\end{equation}
By further differentiating the above, one finds
\begin{equation}
\frac{\partial^3 W}{\partial z^\kappa \partial z^\mu \partial z^\nu}
	(T^{\alpha})^\kappa_\lambda z^\lambda
	+ \frac{\partial^2 W}{\partial z^\kappa \partial z^\nu}
	(T^{\alpha})^\kappa_\mu
	+ \frac{\partial^2 W}{\partial z^\kappa \partial z^\mu}
	(T^{\alpha})^\kappa_\nu = 0.
			\label{ginv3}
\end{equation}
%We fully utilize the above equalities in the following discussions.
The equalities (\ref{ginv2}) and (\ref{ginv3}) become
\begin{eqnarray}
&~& \mu_{AB}=0
			\label{ginv2'}\\
&~& h_{\kappa\mu\nu}(T^{\alpha})^\kappa_\lambda z^\lambda_0
	+ \mu_{\kappa\nu}
	(T^{\alpha})^\kappa_\mu
	+ \mu_{\kappa\mu}
	(T^{\alpha})^\kappa_\nu = O(m_{S})
			\label{ginv3'}
\end{eqnarray}
respectively, in the case of $W$ of Eq. (\ref{W}).
Eq. (\ref{ginv3'}) can be written as
\begin{eqnarray}
&~&h_{AKL}(T^B z_0)^A+\mu_{MK}(T^{B})^M_L+\mu_{ML}(T^{B})^M_K = O(m_{S})
			\label{ginv3'-1}\\
&~&h_{ACL}(T^B z_0)^A+\mu_{MK}(T^{B})^M_C = O(m_{S})
			\label{ginv3'-2}\\
&~&h_{AKk}(T^B z_0)^A+\mu_{MK}(T^{B})^M_k = O(m_{S})
			\label{ginv3'-3}\\
&~&h_{Akl}(T^B z_0)^A = O(m_{S})
			\label{ginv3'-4}\\
&~&h_{ACk}(T^B z_0)^A = O(m_{S})
			\label{ginv3'-5}
\end{eqnarray}
in terms of components.

\newpage

\begin{table}
\footnotesize
\begin{tabular}{lccc}
&$\alpha_i$ & $M_i/\alpha_i$ & $m^2_i$\\ \hline \\
$SU(5) \rightarrow G_{SM}$ & natural & common & testable\\
$SO(10) \rightarrow G_{SM}$ & natural & common & testable\\
$SO(10) \rightarrow G_{PS} \rightarrow
G_{SM}$ & adjustable & common & testable\\
$SO(10) \rightarrow G_{3221} \rightarrow G_{SM}$ & adjustable & common &
not testable\\
$SO(10) \rightarrow G_{3211} \rightarrow G_{SM}$ & adjustable & common &
not testable\\
$SU(5) \times U(1) \rightarrow G_{SM}$ & adjustable & common only for
$i=2,3$ & testable\\
superstring with dilaton $F$-term & adjustable & common & testable\\
superstring with moduli $F$-term & adjustable & not common & not testable
\end{tabular}
\caption[1]{The ``score sheet'' how well we can distinguish between
various models. The intermediate groups are defined as $G_{PS} =
SU(4)_{PS} \times SU(2)_L \times SU(2)_R$, $G_{3221} = SU(3)_C \times
SU(2)_L \times SU(2)_R \times U(1)_{B-L}$, $G_{3211} = SU(3)_C \times
SU(2)_L \times U(1)_R \times U(1)_{B-L}$. The row $\alpha_i$ refers to
the unification of the gauge coupling constants, where ``natural'' means
that the unification is automatic, while ``adjustable'' employs either
particular particle content or threshold corrections to reproduce the
observed gauge coupling constants. The row $M_i/\alpha_i$ refers to the
gaugino masses. The row $m^2_i$ states whether the model predicts
a definite pattern which is testable using the low-energy scalar mass
spectrum. }
\end{table}

\end{document}